\documentclass[preprint,12pt]{elsarticle}
\usepackage{graphicx}
\usepackage{amssymb} 
\usepackage{amsmath}    %needed for \begin{cases}...\end{cases}
\usepackage{latexsym}   %needed for \Box
\usepackage{bm}
\usepackage{times}

\def\half{\textstyle{\frac{1}{2}}}

\begin{document}
  \begin{frontmatter}
    
\title{\vspace*{-1 in}Symbols of a cosmic order}
\author[hm]{F. Hadi Madjid}%% \address[label2]{<address>}
\ead{gmadjid@aol.com}

\author[jm]{John M. Myers\corref{cor1}}
\ead{myers@seas.harvard.edu}

\cortext[cor1]{Corresponding author}
\address[hm]{82 Powers Road, Concord, MA 01742, USA}
\address[jm]{Harvard School of Engineering and Applied  
Sciences, Cambridge, MA 02138, USA}
  
%%%COMMENT OUT ``\today''

\begin{abstract}
The world runs on networks over which signals communicate sequences of
symbols, e.g. numerals.  Examining both engineered and natural
communications networks reveals an unsuspected order that depends on
contact with an unpredictable entity.  This order has three roots.
The first is a proof within quantum theory that no evidence can ever
determine its explanation, so that an agent choosing an explanation
must do so unpredictably.  The second root is the showing that clocks
that step computers do not ``tell time'' but serve as self-adjusting
symbol-handling agents that regulate ``logically synchronized'' motion
in response to unpredictable disturbances.  Such a clock-agent has a
certain independence as well as the capacity to communicate via
unpredictable symbols with other clock-agents and to adjust its own
tick rate in response to that communication.  The third root is the
noticing of unpredictable symbol exchange in natural systems,
including the transmission of symbols found in molecular biology.  We
introduce a {\it symbol-handling agent} as a role played in some cases
by a person, for example a physicist who chooses an explanation of
given experimental outcomes, and in other cases by some other
biological entity, and in still other cases by an inanimate device,
such as a computer-based detector used in physical measurements.
While we forbear to try to explain the propensity of agents at all
levels from cells to civilizations to form and operate networks of
logically synchronized symbol-handling agents, we point to this
propensity as an overlooked cosmic order, an order structured by the
unpredictability ensuing from the proof.  Appreciating the cosmic
order leads to a conception of agency that replaces volition by
unpredictability and reconceives the notion of objectivity in a way
that makes a place for agency in the world as described by physics.
Some specific implications for physics are outlined.
\end{abstract} 

\begin{keyword}
  symbol \sep logical synchronization\sep idiosyncratic phase \sep
  symbol-handling agent \sep Turing machine\sep evidence
  vs. explanation.
\end{keyword}
\end{frontmatter}


%>>>> uncomment following for page numbers
%\pagestyle{plain}    
%>>>> uncomment following to start page numbering at 301 
%\setcounter{page}{301} 
 
%\maketitle

\section{Introduction}
Physicists find numerically expressed regularities in a world that every day
surprises us all with its irregularities.  Recently, commenting on Canales'
book about Bergson and Einstein \cite{canales}, Crease pointed to an
``experiential amnesia'' in physics \cite{crease16}, an amnesia that blocks
attention to pre-conditions for physical time: ``Bergson was trying to bring
to light a sense of time presupposed in the construction of physical time
itself---indeed, in Einstein’s own effort to give to such time a definitive,
mathematical formula.''  The thesis of the present report is that ``time'' as
it works in physics is built out of networks of agent-clocks that do not
dumbly tick, but that self-adjust in response to unpredictable communications
from other clocks of the network.  Examining both engineered and natural
networks of clocks reveals an unsuspected order that depends on contact with
an unpredictable entity. Attention to this unsuspected order has several
implications, including an impact on the notion of scientific objectivity.

Our exposition of this ``cosmic order'' has three roots.  The first
root is the sharpening of a distinction obscured in today's
theoretical physics, namely the distinction between obtaining
numerically expressed evidence from experiments on the laboratory
bench and explaining that evidence in mathematical symbols on the
blackboard.  As reviewed in Sec.\ \ref{sec:2}, the sharpening of the
distinction between physical numerical evidence and numbers calculated
from a theory rests on a proof within the mathematics of quantum
theory that no amount of evidence, represented in quantum theory in
terms of probabilities, can uniquely determine its explanation in
terms of wave functions and linear operators.  Beyond mere opinion,
the proof enables a clarity of thought otherwise unattainable in the
distinction between measured and calculated numbers.  The proof
underpins all the work presented here.  Building on the proof we show
a heretofore overlooked unpredictability of explanations, an
unpredictability beyond quantum uncertainty.  The choice of an
explanation requires an unpredictable reach beyond logic, a fact that
challenges the traditionally notion of objectivity and that precludes
any ``final answers.''
 
The second root stems from our experience with the design of clocking
for fault-tolerant computer networks.  A computer operates one step
after another, regulated by the ticks of its clock.  Fault tolerance
is achieved by using a cluster of several computers, all designed to
do the same task; each computer makes its computational moves in step
with the others, and the computers compare notes at each step.  Their
clocks are organized in a network in which each clock regulates its
tick rate to stay close enough to the other functioning clocks for
comparisons to make sense, but loosely enough so that if one clock
fails, the other clocks continue.  This requires self-adjusting clocks
used not primarily to ``tell time'' but as agents that regulate
motion.  Such a clock-agent has a certain independence as well as the
capacity to communicate with other clock-agents and to adjust its own
tick rate in response to that communication.

Clocks-as-agents are required also by the national and international
organizations that generate time broadcasts.  As spelled out in
\cite{8400_25G}, no two clocks, even those that ``define'' the
international second as a unit of time, tick quite alike.  For this
reason, and because any single clock can fail, the time broadcasts
generated by the National Institutes of Science and Technology (NIST)
depend on several clocks linked by communicated symbols, each clock
adjusting its tick rate so as to receive those symbols at a suitable
phase.  The self-adjustment of clocks requires both computation and
response to unpredictable events, two capacities that might be called
``cognitive''.  In this way our notion of {\it clock-as-agent} has
come to differ rather dramatically from the popular image of ``a
clock.''  In Sec.\ \ref{sec:3} we discuss networks of {\it
  symbol-handling agents} equipped with such clocks, linked pairwise
by communicated symbols, with each agent's clock adjusting its tick
rate so as to receive those symbols at a suitable phase.

The third root is a direction for future research set by noticing
unpredictable symbol exchange in natural systems, including the
transmission of symbols found in molecular biology.

We think of a {\it symbol-handling agent} as a role played in some
cases by a person, for example a physicist who chooses an explanation
of given experimental outcomes or a person in a bucket brigade, and in
other cases by some other biological entity, and in still other cases
by an inanimate device, such as a computer-based detector used in
physical measurements.  We think of a symbol-handling agent as
exhibiting three capabilities:
\begin{enumerate}
\item an agent can transmit and receive symbol-carrying signals to and from
  other agents;
\item an agent can transmit symbols that cannot be predicted prior to their
  transmission;
\item circumstances permitting, an agent maintains a form of
  synchronization---to be called {\it logical synchronization}---with one or
  more other agents, which requires that an agent manage the tick rate of its
  clock.
\end{enumerate}
We will speak of anything that exhibits these three capabilities as a {\it
  symbol-handling agent}, or sometimes, for short, just as {\it an agent}.  (In
earlier work we spoke of symbol-handling agents as {\em live
  clocks}\cite{qip15} or {\em open machines}\cite{aop14})

 As a descriptive form, a network of symbol-handling agents is broad enough to
 encompass the computer network that mediates communications among physicists,
 e.g. the internet with its traffic in both experimental and theoretical
 matters, and also as a form that can be applied to describe naturally arising
 symbolic communication, for example as in the molecular signals of biology.
 Below we discuss the communication of symbols among agents as a neglected
 topic of physics.
 
To model some of the physical behavior of symbol-handling agents linked in
communications networks, we widen an approach pioneered by Turing.  Turing
introduced the Turing machine to model a facet of human cognitive capacity.  In
school we all learn to do sums, to multiply and divide, in short, to
\emph{compute}---a word that in 1936 meant a capability primarily thought of as
a human capability.  In that year Turing abstracted that particular cognitive
capability to produce a specification for an inanimate physical machine.  The
Church-Turing thesis states that a function on the natural numbers is
computable by a human being following an algorithm, ignoring resource
limitations, if and only if it is computable by a Turing machine \cite{CT}.  Since `what
is computable by a human being' is an informal notion, it has no formal
definition, so the thesis, although it has near-universal acceptance, cannot be
formally proven.

In essence, the agent role that we model is that of a physical
sequential processor, that is, a processor that takes one step after
another, with the ``next step'' influenced not just by a stored
program but also by momentary contact with an unpredictable entity.
(Positing that an agent works sequentially answers the question: what
is the difference between one agent and two agents?  Two agents can do
two things concurrently, but one agent can't.)  Our ``agent as a
physical sequential processor'' consists of a self-adjusting clock
that regulates the motion of a Turing machine which is modified to
allow its contact with an unpredictable entity.  For this we use not
the usual Turing machine but Turing's Choice Machine (which we
capitalize for emphasis).  The Choice Machine has the feature, crucial
to our model of an agent, that it can receive symbols from an
unpredictable ``outside''.  Note that the Choice Machine, unlike most
modern practice in computer science, enforces no separation between
program and ``data'', so that symbols received from the unpredictable
``outside'' can work as programs.

Our model of an agent opens a crack in the traditional physicalist image of clockwork
as an explanatory principle in two ways:
\begin{enumerate}
\item By invoking Turing's Choice Machine, we put the `physical sequential
  processor' in contact with an unpredictable entity; and
\item We reformulate the concept of a clock to account for how clocks
  actually work in modern technical contexts, involving communication
  between clocks and their rate adjustments in response to
  unpredictable measured phases.
\end{enumerate}

Years before developing the proof that separates evidence from its
explanations, we experimented on the laboratory bench with conditions under
which the interleaving of two sequences of symbols fails. As described in
Sec.\ \ref{sec:4}, these experiments on instabilities in a decision-making
device (a flip-flop) clarified the circumstances under which two agents could
be expected to agree about the symbol presented to them at a shared moment.
The experimental results on agreement and disagreement contribute to a
re-definition of the notion of ``objectivity.''  In addition they open an
avenue of what might be called ``the physics of borderline cases.''

Section \ref{sec:5} discusses symbol-handling agents in physics as
they appear in several contexts:
\begin{enumerate}
\item The experimental and theoretical working of physicists is
  mirrored in the memories---think Turing tapes---of the computerized
  agents that mediate their communications, e.g. over the internet.
\item Networks of symbol-handling agents, many of them automated,
  serve as tools for experimental inquiry, especially in cases where extreme
  precision is needed, as in the Laser Interferometer Gravitational
  Observatory (LIGO).
\item Networks of symbol-handling agents are promising as metaphors by
  which to describe physical activity, for example in biophysics.
\end{enumerate}
It will be shown that the need for logical synchronization impacts all
these contexts; in particular, in situations requiring the highest attainable
precision of motion, networks of symbol-handling agents cannot derive their
timing as users of national time broadcasts; they have to build their own
``time.''

To reach our conclusions we make the following three assumptions:
\begin{enumerate}
\item We assume we can extend the proposition proven in quantum
  theory concerning the unpredictability of explanations of given
  evidence to physics in general.
\item We partition the (limited) cognitive capabilities of agent that
  we need to be concerned with into computation on one side and
  contact with an unpredictable entity on the other.
\item We assume the Church-Turing thesis.
\item We assume that the communication of sequences of symbols among agents,
  including people, is physical, not in the physics of a clockwork
  automaton, but rather in the physics that we are trying to explicate, a
  physics that has room in it for irreducible unpredictability beyond quantum
  uncertainty.  In particular, we assume that one agent, say Agent A, in
  contact with an unpredictable device such as a photo-detector, can obtain
  from that device a number that a second Agent B, human or not, can find out
  only by the transmission of the number from agent A to B by a signal carrying
  numeric symbols that express the number.
\end{enumerate}
\noindent\textbf{Remarks:}
\begin{enumerate}
  \item We note that the third and fourth assumptions may not be to
    everyone's taste; for alternative, non-reductive views, see
    \cite[Sec. 5.3]{VG}.
\item We do {\it not} offer a theory of cognition; rather we leave
  most of the many capabilities that lurk under the umbrella term {\it
    cognition} undiscussed; we deal only with two: the capability to
  compute, whether by hand or by machine, and the capability to guess
  an explanation, which, as outlined in Sec.\ \ref{sec:2}, requires an
  unpredictable act of a person in contact with an unpredictable
  entity, provably necessary to physics.  We focus on these two
  cognitive capabilities to get at the role of unpredictability
  heretofore overlooked in physics.  Nor do we by any means offer a
  complete characterization of ``agency'', but rather a beachhead into
  a restricted class of symbol-handling agents as descriptive elements
  that have been demonstrated to have interesting applications.  In
  particular we avoid attaching to agency any notion of volition; we
  escape the need to do so by putting in its place an agent's
  responsiveness to something unpredictable.
\item We put a big emphasis on an agent's capacity to issue unpredictable
numeric symbols, and one may ask ``to what do we ascribe the issuance
by an agent of numeric expressions that are unpredictable?''  This is
essentially a question of philosophy or religion, any answer to which
must venture beyond science.  We like to think of the agent issuing
such an expression as being in contact with an unknowable entity, to
which, within the bounds of science, we are precluded from ascribing
any additional features.
\end{enumerate}
While we forbear to try to explain the abundance of behavior at all levels from
cells to civilizations aptly describable as consisting of
logically-synchronized networks of symbol-handling agents in contact with
something unpredictable, in the discussion in Sec.\ \ref{sec:6} we will point to
this abundance as an overlooked cosmic order.

\section{Unpredictability stemming from logically undefined choices in explaining given evidence}\label{sec:2}
We begin this section with a reminder that several varieties of
unpredictability are endemic to physics.  One familiar kind of unpredictability
is the uncertainty that pertains to a spread in a probability measure over a
given set of possible outcomes, implied by quantum theory for a generic
measurement.  Uncertainty makes the outcome at any particular occurrence of a
measurement unpredictable, and in situations that require prompt action,
discussed in Sec.\ \ref{sec:5}, this unpredictability matters.  Another kind of
unpredictability is implied by G\"odel's 1931 proof of the existence of
undecidable propositions in arithmetic, and still another kind by Turing's 1936
proof of the existence of uncomputable functions.  Until fairly recently one
might suppose that uncomputable functions were to be found only in areas of
number theory remote from physics, but now examples are known of undecidable
functions within quantum theory.  Furthermore, as we shall see, Turing's
characterization of computation advanced not just mathematics, but also
physics.

In connection with Turing computability, we note the distinction
between defining a function and computing its values.  For example,
the definition of the square root 2 is distinct from the act of
calculating some number of places of the square root of 2 as a decimal
expression.  The distinction between defining and computing allows one
to speak of certain choices as not merely uncomputable but as
logically undefinable.  Our prime example is the choice of an
explanation of given evidence, as expressed within quantum theory.  We
now review how this choice of an explanation provably requires a reach
beyond logic.

Quantum theory serves as a mathematical language by which to think
about experiments.  Quantum language is built on a skeleton of a
Hilbert space of states and the Born trace rule. This skeleton imparts
a mathematical form to quantum language that remains stable while
hypotheses with physical content come and go.  From this skeleton
alone, we proved in \cite{aop05} that explanations cannot be
determined uniquely from evidence.

The proof can be stated in simplified form, as follows.  To recall the
Born trace rule, we express quantum states as density operators, and
we express measurements by positive operator-valued measures (POVMs).
The simplest POVMs, which are all that we need for a simplified
statement of the proof, assume a finite or a countably infinite set of
outcomes and associate a positive operator $M_i$ to the $i$-th
outcome.  We call these $M_i$ \emph{measurement operators}.  (They sum
to the identity operator: $\sum_iM_i = \bm{1}$.)  In explaining an
experiment in quantum language, one views the experiment as consisting
of a number of trials, each of which involves a preparation followed
by a measurement.  One represents the preparation by a density
operator that can vary from trial to trial.  One expresses the
measurement by a POVM that can also vary from trial to trial.  Given a
density operator $\rho$ and a measurement operator $M_i$, the
probability of outcome $i$ is given by the Born trace rule as
\begin{equation}\label{eq:1}
  \text{tr}(\rho M_i)=\Pr(\text{outcome i})
\end{equation}
The Born trace rule expressed in (\ref{eq:1}) relates probabilities on
the right side of the equation to a density operator and a measurement
operator on the left side of the equation.  An experiment generates
outcomes that can be tallied to approximate probabilities, and the
probability on the right side represents theoretically, on the
blackboard, so to speak, the most one can hope for in the way of
numerical evidence from an experiment.  On the left side of the
equation, the density operator and the measurement operator express an
explanation of the evidence.  Going from left to right in
Eq.\ (\ref{eq:1}), given a density operator and a measurement
operator, the Born trace rule tells how to calculate a unique
probability; many textbook problems ask for just such calculations.
But there are no operators to be seen on the bench, so what happens
when an experiment shows something unpredicted and a new explanation
is needed?  This question presents the inverse problem: given the
experimental outcomes interpreted as probabilities on the right side
of the equation, one seeks a combination of a density operator and a
set of measurement operators that generates the given
probabilities. While the language of quantum theory makes available
density operators and measurement operators as terms to write on the
blackboard, it cannot tell you which density operators and measurement
operators to write in order to explain any particular experiment: for
this inverse problem there is no unique solution.  Given any
probability measure on a countable set of outcomes, we proved that
there are infinitely many POVMs and density operators that generate
the given probability measure. Thus whatever experimental evidence is
on hand admits of numberless explanations.  Choosing one or indeed any
finite number of these explanations requires a reach beyond logic; one
has to guess \cite{aop05}. And a physicist's guessed explanation is
logically unpredictable.

Any experiment can be extended, e.g. by the insertion of another light
source, another detector, another filter, etc.  Correspondingly, any
explanation can be viewed as a restriction to a special case of an
explanation of an extended experiment \cite{tyler07}.  The numberless
explanations that generate a probability measure over a given
parameter domain disagree among themselves with respect to the
extensions from which they can be restricted.  This disagreement
implies that explanations are almost certain to require revision when
tested over enough of their extensions.  The picture emerges of an
open cycle of the making of guessed explanations, testing their
extensions, and revising the guesses; that is, guessing again.\\

\noindent\textbf{Remarks}
\begin{enumerate}
\item The conflicts among predictions of extended explanations are
  \emph{not} limited to small differences associated with issues of
  precision.  In some cases, two explanations that generate given
  probabilities have drastically different implications.  An example
  concerns quantum key distribution, for which two explanations fit a
  given probability measure, but one explanation asserts that the
  quantum key distribution is secure against undetected eavesdropping
  while the other explanation asserts that the key distribution is
  totally insecure \cite{CUP}.
  \item To make a hypothesis in physics is to reach beyond logic to
    assert that a mathematical structure represents certain physical
    behavior.  Because of its inescapable dependence on guesswork, the
    hypothesis is subject to eventual falsification.  It is worth
    noting that, unlike a hypothesis stated in the language of quantum
    theory, the language itself is \emph{not} falsifiable: given any
    probability measure on a countable set of outcomes, there exist
    (many) quantum explanations, so there can be no probabilities that
    lie outside the reach of quantum language \cite{CUP}.  Being
    unfalsifiable, the language of quantum theory offers relatively
    stable mathematics.  The proof of the need for guesswork to choose
    an explanation is not a hypothesis of physics, but a proof within
    mathematics.
\item Coming up with an explanation necessarily involves an act of
  imagination beyond anything calculable.  An act of imagination
  requires intimate contact with something unknowable. (If you know
  it, you don't have to imagine it.)  The finding that explanations
  cannot be calculated from data but require contact with something
  unknowable has the following implication that adds to the necessity
  in physics for communication.  An person's act of imagination
  expressed in symbols remains unknowable to a second person unless
  the symbols of the expression are communicated from the person to
  the second person.
\item Recognizing guesswork and its vulnerability to occasions for
  revision as indispensable to explanations precludes the possibility
  of assuring any final answer or any ``objective truth.'' Yet, under
  circumstances to be discussed in the next section, agents can
  communicate sequences of numerals expressing evidence and
  explanations, numerals about which they can be expected to agree;
  remarkable too is that the actions needed to maintain the
  communications channels depend on idiosyncratic contact with
  something unknowable, beyond collective agreement.
\end{enumerate}

\section{Networks symbol-handling agents}\label{sec:3}
This section is dense with detail needed for the design of experiments
aimed at exploring the cosmic order.  In it we give modeling language
to express agents that recognize symbols carried by signals. By
\emph{symbols}, we mean elements of communication among agents, human
or not, that can lead to action involving energy supplied not by the
symbols themselves nor by the agent that sends them but by the
receiving agent.  How a receiving agent responds to a symbol depends
on the past history of the agent.  Symbols can be conveyed from agent
to agent in various ways, for example by: (1) letters, words, and
numerals, expressed as written characters, and (2) electronic impulses
carrying bits in a computer, and (3) molecules involved in biological
signaling. Symbols used by people convey elements of thought, often
prompting actions and emotions.  And symbols convey the calculational
traffic to be found both in man-made computers and in the biological
processes of living creatures.

Whether to say \emph{signal} or \emph{symbol} as we use these words is
a matter of point of view, and not a matter of distinguishing two
distinct things.  Consider a pawn on a chess board.  From the
viewpoint of players, the pawn is a symbol.  A player is indifferent
to variations within certain bounds of its shape, its weight, and when
and how it is placed on a square of the chessboard, e.g. whether it is
a little off center doesn't matter.  For the craftsman who makes the
pawn, however, its signal aspect matters: its shape, its weight, the
material of which it is made.  The same holds for the 0's and 1's in a
computer.  To a programmer they are symbols, while to an electrical
circuit designer the details of the physical signals that the
programmer views as symbols matter greatly. It is a fact, indeed to us
an amazing fact, that symbols embodied by signals with large
tolerances can take part in mechanisms that act with tight
tolerances. This is what happens in computer-controlled machining: the
signals that carry the 0s and 1s in the computer can vary within broad
limits without interfering with the mathematical exactitude of the
calculations that control the shapes that are machined.

Numeric symbols arrive one after another, and the sequence matters:
``1011'' is a different message from ``1110''.  An agent must deal
with symbols in a way that respects the order in which they arrive,
making it convenient to view an agent as a sequential processor.
Turing broke new ground by expressing a sequential processor
mathematically by his ``machines'', both what today is called a Turing
machine and a lesser known Choice Machine that plays the crucial role
in our model of an agent.

Offering mathematical structure to represent physical behavior is the
business of physics, and in his ``machines,'' Turing made an
interpretive leap beyond the reach of mathematics to create novel
physics.  In representing the physical activity of computing, he did
not limit himself the terms (e.g. ``particles and fields'') expected
in physics; rather, he introduced new terms (e.g `tape', 'scanned
square', `moment', `move') peculiarly appropriate to describing the
activity of a calculational agent.  In retrospect, one also sees
physics implicit in G\"odel's 1931 proof of undecidability, because
proving a proposition is a physical activity: one has to write into a
recording medium, and in checking a claimed proof, one reads the
written record.

In the rest of this section we consider agents who communicate by
means of transmitted sequences of symbols from one agent to
another. We aim to show the critical role of the timing required.  To
this end we assume that each agent is a sequential processor.  As a
model of an agent that communicates with others, we start with a
Turing machine as a model of a sequential processor, and then modify
it, first by giving the machine the capacity to transmit and to
receive symbols from other such machines, and secondly by modifying
the Turing machine to be stepped by an accompanying clock that can
have its rate adjusted by commands from the Turing machine.

The Turing machine as a sequential processor is elegantly described
in Turing's 1936 paper: 
\begin{quote}
  We may compare a man in the process of computing a real number to a
  machine which is only capable of a finite number of conditions $q_1,
  q_2, \ldots, q_R$ which will be called ``$m$-configurations''. The
  machine is supplied with a ``tape'' (the analogue of paper) running
  through it, and divided into sections (called ``squares'') each
  capable of bearing a ``symbol''. At any moment there is just one
  square, say the $r$-th, bearing the symbol $S(r)$ which is ``in the
  machine''. We may call this square the ``scanned square''. The
  symbol on the scanned square may be called the ``scanned symbol''.
  The ``scanned symbol'' is the only one of which the machine is, so
  to speak, ``directly aware''. However, by altering its
  $m$-configuration the machine can effectively remember some of the
  symbols which it has ``seen'' (scanned) previously. The possible
  behaviour of the machine at any moment is determined by the
  $m$-configuration $q_n$ and the scanned symbol $S(r)$. This pair
  $q_n$, $S(r)$ will be called the “configuration”: thus the
  configuration determines the possible behaviour of the machine. In
  some of the configurations in which the scanned square is blank
  (i.e. bears no symbol) the machine writes down a new symbol on the
  scanned square: in other configurations it erases the scanned
  symbol. The machine may also change the square which is being
  scanned, but only by shifting it one place to right or left. In
  addition to any of these operations the $m$-configuration may be
  changed.  Some of the symbols written down will form the sequence of
  figures which is the decimal of the real number which is being
  computed. The others are just rough notes to ``assist the
  memory''. It will only be these rough notes which will be liable to
  erasure.\cite{turing}
\end{quote}
In a side remark in the same paper, Turing briefly introduced an
alternative machine called a {\em choice machine}, contrasted with the
usual Turing machine that Turing called an a-machine:
\begin{quote}
  If at each stage the motion of a machine \ldots is completely
  determined by the configuration, we shall call the machine an
  ``automatic machine'' (or a-machine). For some purposes we might use
  machines (choice machines or c-machines) whose motion is only
  partially determined by the configuration \dots. When such a machine
  reaches one of these ambiguous configurations, it cannot go on until
  some arbitrary choice has been made by an external operator. This
  would be the case if we were using machines to deal with axiomatic
  systems \cite{turing}.
\end{quote}
One of the two ingredients in our model of a symbol-handling agent is
a Choice Machine modified so that it can take part in a communications
network by transmitting symbols to other such machines and, under
certain conditions of "logical synchronization," can receive symbols
transmitted to it by other machines.  To assure the unpredictability
of a symbol-handling agent, we posit that an ``external operator''
chooses a symbol and writes it onto the scanned square of the agent's
Choice Machine {\em privately}, in the sense that the symbol remains
unknown to other agents unless and until the symbol-handling agent
that receives the chosen symbol reports it to others.

\subsection{The motion of symbol-handling agents.}
Sequences of symbols on the squares of a tape at a moment are
static---they show no motion.  To deal with the physics of motion, one
needs to deal not just with recorded sequences but also with temporal
sequences, such as sheep herded one after another through a gate or
the back-and-forth swings of a pendulum.

To express motion mathematically, one faces the hurdle that what a
mathematical function does is to assign an element of a codomain
to each element of a domain, with both the domain and the codomain
thought of as static.  From the standpoint of physics, motion
expressed mathematically is essentially motion captured in the frames
of a movie film detached from information about scale and speed.  The
movie film would be the same if the scene filmed were sped up by a
factor and the clock that drives the movie camera were sped up by the
same factor, so that in mathematics one cannot express motion
\textit{per se}.  In preparation for discussing the regulation of the
motion of agents necessary to their exchange of symbols, we flag two
points:
\begin{enumerate}
\item In the mathematics of the Choice Machine, a `move' is a mapping
  of a machine configuration at one moment to a machine configuration
  at the next moment, and a sequence of moments is like a sequence of
  frames of a movie film, which, by itself, contains no specification
  of its physical motion, for example, how rapidly it is to be moved.
\item Implicit in the interpretation of the Choice Machine as
  representing the physical action of computation is the subdivision
  of the moment into a phase in which the machine can read the scanned
  square, so that its next action is determined, and a distinct other
  phase in which the machine can write on the square.  This
  subdivision is necessary to avoid a conflict between trying to read
  and trying to write on the same square at once.
\end{enumerate}
With these points in mind, we think of a
Choice Machine moved by the ticks of a clock at a rate adjustable
(relative to the clock's own unadjusted internal standard) by commands
from the Turing machine itself, as discussed in \cite{qip15,aop14}.
If we picture the clock as having a single hand that cycles around a
dial, then subdivisions of the dial correspond to phases of the
computational cycle of the Turing machine, with a phase in which a
symbol can be written on the scanned square.  This modified Choice
Machine expresses a computer that can take part in a communications
network, and as such combines both the logic and the motion required
of a process-control computer in contact with an unpredictable
environment.  The modified Choice Machine stepped by its adjustable
ticks models our {\it symbol-handling agent}.

\subsection{Logical synchronization.}
We call the condition in which symbols arrive during a phase of the
computational cycle in which they can be written into memory
\emph{logical synchronization}.  The need for logical synchronization,
long known to engineers of digital communications \cite{meyr,meyr2}, is
reminiscent of a game of catch, in which a player cycles through
phases of throwing and catching a ball, or perhaps more simply a
spoken dialog in which each person alternates between listening and
speaking.  The requirement for logical synchronization constrains
number-carrying networks. (If a spacetime manifold in invoked, logical
synchronization put ``stripes'' on spacetime \cite{aop14}.)

Another image of logically synchronized symbol-handling is a bucket
brigade, in which people work in a line, each passing a full bucket to
the person to the right while receiving an empty bucket in exchange,
which is next passed to the person to the left.  The people in a
bucket brigade work in dovetailing cycles.  Each cycle contains a
phase in which one person passes a full bucket to a neighboring
person, sharing a rhythm: if you are to my right, then when you turn
to the left, I turn to the right to pass my full bucket to you and to
receive your empty bucket.  If I try to pass you my full bucket
outside of the phase in which you can receive it, we spill the bucket.

Logical synchronization requires more or less continual adjustment of
clock rates to limit the drift of physical clocks.  Unavoidable drift
in clock rates stems from quantum uncertainty, from relative motion of
the agents, and from other causes \cite{aop14}.  The adjustment of
tick rates of agents' clocks entails feedback that responds to the
phases at which transmitted symbols arrive.  On the blackboard, we
represent the cycle of the agent's clock by a unit interval of the
readings of its adjustable clock, and we express a reading of a the
clock as $m.\phi_m$ where an integer $m$ indicates a count of cycles
and $\phi_m$ is the phase within the cycle.  Choosing the convention
that $-1/2<\phi_m\le 1/2$, we model the phase of writing at which an
agent can receive a character as corresponding to
\begin{equation}\label{eq:ls}
  |\phi| < (1-\eta)/2,
\end{equation}
where $\eta$ (with 0 $< \eta < 1$) is a phase interval that makes room for
reading.  When this phase constraint is met for a channel between a
transmitting agent and a receiving agent, we say the receiving agent is
\emph{logically synchronized} to the transmitting agent.

The adjustment of the rate of an agent's clock in order to maintain
logical synchronization with another agent proceeds as a balancing
operation.  We think of an agent's clock as having a ``faster-slower
lever.''  The ``faster-slower lever'' works like the pointer of a
balance instrument that moves one way or the other in response to
opposing impulses from an arriving symbol and from the agent's
reference signal in the middle of the receptive phase.  If the symbol
arrives in the middle of the receptive phase, neither well before nor
well after the reference signal, the response of the balance is
indeterminate; it can tip a little either way or hover in the middle.
While symbol recognition is invariant under limited variations in
timing, making it also invariant under interchange of manufactured
instruments, the balancing that drives rate adjustment will not be the
same if the balancing instrument is interchanged with another of the
same manufacture. We speak of the behavior of measured phases that
vary when two balancing components of the same design are interchanged
as \emph{idiosyncratic}.

Logical synchronization depends on the happy fact that the
idiosyncrasy endemic to balancing does not matter, because it occurs
only when the symbol arrives more or less in the middle of the
receptive phase, so that a small adjustment of clock rate cannot cause
the next few symbols to arrive outside their receptive phases.  If the
agent's clock with its ambiguous small adjustments drifts enough so
that the subsequent arrivals are noticeably early or late relative to
the mid-phase aiming point, the operation of the balance becomes
definite and adequately corrective.

\begin{quote}
\noindent\textbf{Remark:} Maintaining logical synchronization depends
on prompt steering in response to deviations that can be expressed
numerically only later, after they have been responded to.
Furthermore, the numerical expression of the deviations is necessarily
idiosyncratic in that neither two persons nor two machines can be
expected to produce numerical records of phases that agree.
\end{quote}

To represent the transmission of numerals from one agent to another on
the blackboard, we follow Shannon in speaking of a communications
channel; however we augment his information-theoretic concept of a
channel \cite{shannon48} with the agent's clock readings at the
transmission and reception of symbol-bearing signals \cite{aop14}.  We
indicate the timing in a \emph{channel} from agent $A$ to agent $B$,
by a set of pairs, each pair of the form $(m.\phi_m,n.\phi_n)$.  The
first member $m.\phi_m$ is an $A$-reading at which agent $A$ transmits
a signal and the second member $n.\phi_n$ is a $B$-reading at which
agent $B$ registers the reception of the signal.  In this way the
notion of a channel is expanded to include the clock readings that
indicate phases of signal arrivals that have to be controlled in order
for the logical synchronization of the channel to be maintained.
\begin{quote}
  \textbf{Proposition:} A symbol can propagate from one agent to
  another only if the symbol arrives within the writing phase of the
  receiving agent.

  \noindent\textbf{Corollary:} Exact agreement concerning symbols
  depends on idiosyncratic management of phases.
\end{quote}

\section{Sequencing failures and the concept of objectivity} \label{sec:4}
What happens when a symbol-carrying signal arrives at a receiving agent just
too late for the agent's writing phase?  Forty years ago we investigated this
question experimentally by examining the behavior of an elemental decision
agent, a clocked flip-flop.  Packed by the million on the silicon chips of
communicating computers, the clocked flip-flop is a memory device into which
a 0 or 1 can be written, provided that the symbol arrives during a phase in
which the clock opens a gate.  A flip-flop is the electronic analog of a
hinge that records a 1 if flipped one way or a 0 if flopped the other way.
In the case of an electronic flip-flop, an arriving 1 comes embodied as an
electrical pulse of energy above a high threshold, and an arriving 0 comes
with an energy below a low threshold, well under the high threshold.  For a
flip-flop set at 0, a 1 arriving while the gate opens on a phase of writing
can flip the hinge over to record a 1.  In effect, the clocked flip-flop is
aware of the possible symbols that it might receive, in that it balances any
arriving signal against a reference energy level. If the pulse energy is
above the reference level, the flip-flop both augments the pulse energy to
lift it above the high threshold and flips the hinge over. If the pulse
energy is below the reference level, the flip-flop drains its energy below
the lower threshold and stays flopped back to indicate a 0.

A symbol arriving after the clock closes the gate is shut out and
ignored by the flip-flop; however, if a pulse of electrical energy
conveying a 1 arrives just as the writing phase is ending, in a race
with the clock's closing of the gate, the pulse squeaks through the
closing gate into the receiving flip-flop as a ``runt pulse.''  Then
the hinge might flip to a 1 or might stay at 0, but there is another
possibility---indeed a possibility outside the frame in which digital
signals are conventionally discussed: the runt pulse can be so close
to the reference level that it lifts the hinge part way but not all
the way over, leaving the hinge hung up in an unstable ``in between''
state, teetering on edge, until, eventually, it flips or flops
\cite{glitch,glitch2,glitch3}.

This ``in between'' state occurs occasionally when a computer responds
to unsynchronized input signals, and it leads to logical confusion, as
follows.  Computations require that a receiving clocked flip-flop $A$
transmit copies of its record not just to one following flip-flop but
through a fan-out to a pair of flip-flops, say $B$ and $C$, so that
whatever symbol was in $A$ at an earlier moment appears in both $B$
and $C$ at the next moment; that is, $B$ and $C$ both receive 0 or
both receive 1.  But if flip-flop $A$ teeters in an unstable
equilibrium, then flip-flops $B$ and $C$ may not only hang up, but can
``fall differently'' so that the symbol in $B$, instead of matching
that in $C$, conflicts with it, which is what we mean by ``logical
confusion.''

We explored the teetering of a flip-flop not just analytically on the
blackboard, but as it takes place physically on an electronics bench.  The
results, reported in \cite{aop05}, were clear enough, but we could not
describe what was to us the most interesting aspects of the experiment; words
failed.  With our subsequent proof that puts unpredictability squarely within
physics, however, we can now tell the story.  In particular, the experimental
design to be described illustrates how the investigators and the laboratory
bench shape each other, how the investigators must be aware of the need to
adjust the devices on the bench, how the devices must be designed to be
receptive to adjustment by the investigator, how unpredictable outcomes
propagate from the bench to the investigator, and how the unpredictable
responses of the investigator to these outcomes propagate back to the bench.

Our experiment to measure the teetering of a clocked flip-flop consisted of
several billion trials of sending a 1-pulse to a clocked flip flop $A$ that,
after an adjustable delay $T$, was read by two flip-flops $B$ and $C$.  It
took about 300 nanoseconds per trial.  For each trial we arranged for the
1-pulse to race a gate-closing clock pulse, causing a runt pulse that made
the flip-flop $A$ teeter on edge.  To tell if $A$ was teetering on edge, we
arranged the electronics to keep a running average over past trials to record
how often the flip-flop, after teetering, fell to 1 vs. how often it fell to
zero.  If the running average was about even between 1's to 0's, then the
timing of the of the 1-pulse relative to the clock pulse was likely to make
the flip-flop teeter.

But when we first tried to set the timing so as to bring about this desired
even running average, we ran into trouble.  We generated both the clock pulse
and the 1-pulse repetitively at 300 ns intervals through a delay line
from a common signal generator, and we regulated the timing of the 1-pulse
relative to the clock pulse by adjusting the delay of the delay line for the
1-pulse.  The trouble was that however we adjusted the delay line, we got
either all 1's or all 0's from the running average.  Because of drift in the
delay line and perhaps other unanalyzed effects, we could not get the very
delicate balancing of the ``hinge'' that we wanted.  The solution was to use
feedback.  We electronically read the running average and fed that back to
automatically adjust the delay line.  Feedback worked like a charm to put the
flip-flop on edge well enough to run the experiment.

Another interesting feature of the experimental design was the detection of
teetering.  Prior investigations of teetering in flip-flops made use of
oscilloscopes that operated on faster time scales than that of the flip-flop;
in effect the oscilloscope acts as a movie camera to photograph the teetering
of the (electronic) hinge.  Instead of using an external high-speed
``camera'', we wanted an experimental design that works even if the flip-flop
operates faster than any such external ``camera.'' To this end, instead of an
oscilloscope, we used the logical confusion ensuing from a runt pulse as a
means of measurement, by using the pair of flip-flops $B$ and $C$, as
described above.  We could then plot the statistics of how often $B$ and $C$
disagreed with each other as a function of the waiting time $T$, as reported
in \cite{aop05}.

Feedback, though it worked like a charm, posed what then seemed to us
to be a conundrum. We were used to thinking of an experiment as
something that an investigator may start, but must then keep his or
her hands off until a result emerges.  With feedback, we put
ourselves, or at least the automated feedback loop that acted for us,
right into the operation of each trial, so that we were using outcomes
from earlier trials to direct a current trial.  Now we say: we as
investigators act like symbol-handling agents, and that's what
symbol-handling agents do.  But still, the acceptance of feedback
invites one to rethink the conventional devotion to ``objectivity'' as
a Cartesian preclusion of the observer from tinkering with the
observed.  By admitting feedback from previous trials, one can explore
unstable physical behavior not otherwise susceptible to investigation.

In connection with ``objectivity'' there is something interesting
about the use the pair of receiving flip-flops $B$ and $C$ to detect
teetering. When flip-flop $A$ is receiving a 1-symbol under conditions
of logical synchronization, so the 1-symbol arrives well within the
phase of writing, there is no disagreement between $B$ and $C$.  The
flip-flops $B$ and $C$ act as agents that are interchangeable.  Such
interchangeability is a hallmark of what we view as a reconceived
``objectivity'': over a sequence of trials in which $A$ is sent at one
moment sometimes a 0 and sometimes a 1, at the following moment $B$
and $C$ reliably show the same outcome: both show 0 or both show 1.
The discrepancies registered in the experiment show that the
measurement of a phase $\phi$ cannot be objective in the way we mean
it: two detecting agents can disagree; indeed we used what might be
called the \emph{idiosyncrasy of the agents $B$ and $C$} to indicate
the phase corresponding to the closing of the gate.  As we said in the
previous section, measurements of phase, necessarily idiosyncratic,
are indispensable to the logical synchronization that enables
interchangeable agents to agree about counts of cycles and sequences
of symbols received.

In summary, by introducing a distinction between symbols registered
interchangeably by several agents and the idiosyncratic phases that must be
responded to in order to maintain logical synchronization, the investigation of
a race between a symbol and a clock tick exhibits a crack in the Cartesian
veneer of disembodied objectivity in which theoretical physics pretends to wrap
we who investigate:
\begin{enumerate}
\item The race condition could not be passively observed but had to be actively
  maintained in the face of disruptive effects that we could not predict.
\item Maintenance of the race condition depends on the investigator choosing
  a control algorithm that specifies how a feedback loop responds to
  unpredicted effects, a choice that requires an act of imagination.
\item The most sensitive indicator of a race condition is a ``measurable breakdown in
  Cartesian objectivity'': two detectors disagree.
\end{enumerate}

\section{Unpredictable, symbol-handling agents in physics}\label{sec:5}

\subsection{Physicists communicate with each  other and with their experimental devices.}\label{sec:5A}

Physics depends on the transmission of symbols, such as the sequences
of symbols that make numerical records, both records of experimental
results and records of calculations. Yet until now theoretical
physics, with its emphasis on particles and fields, has had no place
in its vocabulary for {\it record} or {\it symbol}.  The absence of
{\it symbol} from the vocabulary was perhaps appropriate under the
out-of-date picture of an experiment as producing a stream of records
in which no record depends on the records produced earlier.  But
today's experiments involve computer-mediated feedback, in which
symbols conveying theoretical calculations and symbols conveying
experimental results combine to update calculations and to steer the
experiment.  The physical behaviors possible as targets of
investigation depend on the possibilities for the transmission of
symbols, so that it no longer makes sense to leave {\it symbol} and
{\it record} out of the vocabulary of theoretical physics.

With regard to the application of theory to physical situations,
one thinks more clearly by picturing numeric symbols expressing both
calculations and measured outcomes as resident in agents' memories,
e.g. as on their tapes.  Thus we see the symbols of (\ref{eq:1})
as written ``on tape'', in sharp distinction to any physical evidence.  From
Sec.\ \ref{sec:2} it follows that the symbols of an explanation can
never by uniquely determined by evidence, so that a written
explanation depends on a logically unpredictable choice of a theorist.

Once one sees explanatory statements this way, one is free to straddle
different theoretical frameworks (rationalizing what is done anyway):
parts of any big experiment, such as those conducted at the Large
Hadron Collider, are described in essentially Newtonian terms, while
other parts are described in the language of quantum theory.  One is
free to think in whatever mode one chooses for one or another part or
aspect of an explanation.  For example, in \cite{aop05}, we offer a
quantum-theoretic explanation, indeed two of them, for the flip-flop,
a device that is also conveniently explained in terms of classical
circuit theory.

\subsection{Regulation and measurement of motion.}
Traditional physics presupposes coordinate systems as mathematical
constructions that one relates to physical systems.  Coordinate
systems are defined (at least locally) by Einstein's imagined patterns
of light signals propagating between imagined \emph{proper clocks}.
In terms of proper clocks and signals, Einstein defined the
synchronization of proper clocks fixed to a non-rotating, rigid body
in free fall (i.e., a Lorentz frame), and he co-defined ``time'' as the
readings of such proper clocks, with the implication that distance
from proper clock $A$ to proper clock $B$ is defined, as in radar, in
terms of the duration at $A$ from the transmission of a light signal
to the return of its echo from $B$.  Specifically, according to
Einstein's definition of the synchronization of proper clocks
\cite{einstein05}, clock $B$ is \emph{synchronous} to clock $A$ if at
any $A$-reading $t_A$, $A$ could send a signal reaching $B$ at
$B$-reading $t_B$, such that an echo from $B$ would reach $A$ at
$A$-reading $t'_A$, satisfying the criterion
\begin{equation}\label{eq:es}
t_B=\half(t_A+t'_A).  
\end{equation}

Unlike logical synchronization with its explicit dependence on
idiosyncratic responses to phases, Einstein's synchronization criteria
are blackboard criteria that take no account of the responses to
phases necessary to the communication of symbols. To implement a
coordinate frame, actual signals conveying numerals as symbols are
necessary. In \cite{qip15,aop14} we tell how, in a generic curved
spacetime, there is no dense set of clocks that can pairwise satisfy
Einstein's synchronization criterion, but a finite network of
symbol-handling agents can act as a reference frame; furthermore, such
a network can serve as a detector of gravitational radiation.  In
\cite{qip15} we also note that clocks on a rotating platform (such as
a merry-go-round) can never satisfy Einstein's synchronization
criteria but still can be logically synchronized.

In practice, computerized signal-handling agents take part in the
generation of Universal Coordinated Time (UTC), as well as in the
operation of the Global Positioning System (GPS). As noted above, to
generate time broadcasts, NIST must adjust the rates at which its
clocks tick in relation to one another.  This adjustment depends on
the communication of numeric symbols in a system of feedback loops
\cite{aop14}.

Besides noticing that time broadcasts depend on networks of agents
that respond to unpredictable deviations, we call attention to the
freedom of investigators to ``build their own time'', tailored to
their particular investigations.  This is routine in experiments
investigating the instability of cutting-edge optical clocks: one
achieves much higher precision by comparing one clock directly against
another; the idea of comparing each to NIST time makes no sense, for
the best optical clocks have much smaller instabilities than does NIST
time.  Similar direct comparison that bypass ``time broadcasts'' are
required for LIGO.

\subsection{Occurrences of outcomes used promptly in feedback.}
Quantum-theoretic models predict probabilities of outcomes but the
occurrence or non-occurrence of an outcome at a particular trial of a
generic measurement is, by postulate, unpredictable.  In experiments
without feedback, the occurrences of outcomes over a run of trials are
tallied up but not otherwise acted on, a fact that can obscure the
significance of the unpredictability of occurrences of outcomes.  With
feedback in which an agent responds promptly to one or a few
occurrences of outcomes to bring about physical behavior not otherwise
attainable, as discussed above, one sees essential unpredictability at
work, not just in the mind of the physicist, but also on the workbench
of experiments, for example in a photo-detector that may or may not
respond to light at the single-photon level.

\subsection{Changes in principle brought by the recognition of symbol-handling agents.}
Timing controlled by feedback stands in contrast to ``time'' as a
concept in physics, whether the ``time'' is that of Newton or that of
Einstein, for both concepts of ``time,'' stand outside of the whatever
is under investigation.  As Newton put it:
\begin{quote}
 Absolute, true, and mathematical time, in and of itself and of its
 own nature, without reference to anything external, flows uniformly
 and by another name is called duration.  Relative, apparent, and
 common time is any sensible and external measure (exact or
 nonuniform) of duration by means of motion; such a measure---for
 example, an hour, a day, a month, a year---is commonly used instead
 of true time \cite{newton_t}.
\end{quote}
While Newton says that ``true time'' does not refer to anything
external, this is a fudge, because he postulates ``true time'' as
externally provided, independent of what any person or other entity of
interest does.  Two centuries after Newton, Einstein made time
relative to the concept of a proper clock, but the proper clock is
imagined to tick at a rate that is again externally provided. And
neither in Newtonian physics nor in special or general relativity is
the distinction drawn between evidence on the laboratory bench and the
formulas written on a blackboard, making it a challenge to think in
terms of this distinction; yet the distinction between evidence and
its explanations must be made in coming to grips with the role of
symbols in physics.

\subsubsection{Experimental freedom to set aside the assumption of spacetime.}
The assumption of a spacetime manifold as an explanatory principle has
had a dominant place in physics for decades; these days, however, from
a number of sources, one would like to be able to set aside that
assumption.  An issue is that the assumption of a spacetime manifold
is built into the reference system which the International
Astronomical Union (IAU) offers for the location of events, such as
the event of the tick of an agent's clock \cite{soffel03}.  Although
for many purposes convenient, the assumption of a spacetime manifold
in an unnecessary impediment to exploration.  As shown in
\cite{qip15}, the records of a network of agents of clock readings at
the transmission and the reception of signals form a basis, free of
any assumption of a manifold, against which to experimentally test
hypotheses of spacetime manifolds.

\subsubsection{Limitations on the interleaving of sequences.}
Consider the case of a three-way race among signals $X$, $Y$, and $Z$
arriving at a place at which they are to be temporally ordered. Such a
comparison involves pairwise balancing, involving flip-flops or their
equivalent as decision elements.  Each of the three signals fans out
to allow three separate pairwise comparisons of “which came before
which”. In a close race, teetering in all three pairwise comparisons
can result in finding: $X$ before $Y$, $Y$ before $Z$, and $Z$ before
$X$, rather than the ``expected'' $X$ before $Z$, violating the
transitivity of an ordering relation, and suggesting a limit on the
validity of even local temporal ordering. Making sense out of temporal
order requires distinguishing the question of which cycle a symbol
recognition occurred from the question of the phase of a cycle at
which a signal arrived \cite{1639}.

\subsubsection{Irreversibility of unpredictable events.}
The basic equations of physics involve a time variable $t$ and are
invariant under the transformation $t \rightarrow -t$.  When the
mathematical language of experimental physics is over-stressed, this
invariance appears to impose time reversibility as a principle of
physics, in conflict with thermodynamics. But, as we have emphasized,
equations written on the blackboard are not the whole story in
physics.  Both because of their unpredictable choices and their action
in feedback loops that respond to unpredictable occurrences of
outcomes, symbol-handling agents introduce a heretofore overlooked
source of irreversibility into physics, even when the equations they
write on the blackboard are invariant under $t \rightarrow -t$.  As we
explain in \cite{spie2016}, widening the scope of descriptions
admissible to physics to include the agents and the symbols that link
theory to experiments opens up a new source of time-irreversibility in
physics.

\subsection{Networks of symbol-handling agents as metaphors.}
And beyond the role of symbols in time broadcasts and in
investigations conducted by physicists, symbols carry information in
networks other than those used by people, including networks in active
matter in the living world, for example, as codes written in the
nucleobases of strands of DNA molecules, or the sequences of
amino-acid side chains of a protein molecule.

\section{Discussion} \label{sec:6}
The first root of the recognition of symbol-handling agents came from
the proof that sharpened the separation between evidence and its
explanations.  Our awareness of this separation comes and goes; this
awareness is a limited resource, hardly to be maintained in the midst
of calculations, so one can have no ``once-and-for-all'' separation;
however; we can and do find occasions to separate our thinking at the
blackboard from our thoughts about the bench to which we apply the
blackboard, and sometimes this turns out to be very productive. An
example to do with cryptography is mentioned at the end of
\cite{aop14}.  When we recognize that our connecting of an explanation
to evidence takes an act of our own guesswork, we recognize our own
agency in the physical world.

The recognition of our own agency, our own participation in guesswork,
influenced by our own individuality, opposes a long history of efforts
to claim for physics an ``objectivity'' that goes back to the
scientific outlook of Descartes.  As Riskin summarizes Descartes's
stance:
\begin{quote}
  Seeing the world as pure machine, lifting his thinking soul out of the world,
  even out of its own bodily interface with the world, Descartes accomplished
  the distancing of self from world that defines modern subjectivity, the sense
  of a fully autonomous, inner selfhood, and modern objectivity, the sense of
  regarding the world from a neutral position outside of it \cite{riskin}.
\end{quote}
In Sec.\ \ref{sec:2} we reviewed the separation of the blackboard of
theory from evidence on the work bench, leading to the necessity of a
guess beyond the reach of logic to bridge that separation.  It is a
guess that selects an explanation of evidence, or even that narrows
the selection to any finite number of explanations.  Because
explanations enter both the design and the operation of experiments
that generate evidence, the unavoidable guess makes a place in physics
for the personal acts of imagination of physicists.

The proven dependence of physics on guesses as acts of imagination
refutes any claim of a quantum explanation to an ``objectivity''
that aspires to produce a ``final truth'' from a neutral position
outside the world investigated. As an investigator, I work with
guessed assumptions, some of which I change from time to time.  I
climb about about on a ``tree of assumptions,'' able, perhaps, to let
go of this or that assumption, but only by taking hold of other
assumptions \cite{aop05}; there is no way for me (or for you) to look
at the tree of assumptions from outside it.  Thus ``objectivity'' as a
neutral position outside the world investigated makes no sense even
as a goal, but a different, less global, notion of objectivity
survives, in that, under appropriate circumstances, logical
communication that two agents can agree about remains possible.  By
way of illustration, after the experiments on the reception of signals
discussed in Sec.\ \ref{sec:4}, we did some experiments on people
counting. As written in notes of JMM:
\begin{quote}
In 1982, I asked my son Sam and his friend Gordon to act as agents in
some experiments on counting.  I put a few paper cups on an otherwise
bare table and asked them to each write down on a slip of paper the
number of cups.  In one trial they looked at the table together, in
another trial they entered the room one after another; in both cases,
unsurprisingly, they counted the same number of cups.  Then we did
another series of trials in which the two boys viewed the same table
at the same time, as told by a second hand on wall clock that they
could both see.  At each trial I first cleared the table and then gave
the boys a starting signal some seconds before the second hand crossed
the twelve o'clock mark.  While they watched the table I would put
paper cups on it, and in some trials I would slide some cups and
remove some while adding others. Their job was to each separately
write down the number of cups on the tables as the clock hand passed
12.  While they would get the same number if the cups were not in
motion, when I fiddled with the cups while they counted them, they
usually wrote down counts that were in disagreement, e.g one wrote
down ``8'' while the other wrote down ``11.''
\end{quote}
From that experiment come two interesting results.  One was bringing
down to the work bench, in this case the table, the notion of an
objective number as a number reported by interchangeable agents, and
so independent of which boy reported it.  The other was the
distinction between the situation in which change was blocked while a
count was made---in effect logical synchronization---and the situation
which, as in the experiments on logical confusion, the moving and the
viewing were unsynchronized, leading to disagreement.  We find then
that objective counts are indeed possible but only under circumstances
of logical synchronization.

While we forbear to try to explain the empirical propensity of agents at all
levels from cells to civilizations to form and operate networks of logically
synchronized symbolic communication, we point to this propensity as an
overlooked cosmic order, ripe for further investigation.

\section*{Acknowledgment}
Our thanks to Samuel Myers and Gordon Burnes for taking part in counting
experiments, and to Jessica Riskin for an exchange of emails concerning the
historical pulling and tugging over the role in physics, or lack of it, for an
agent.  We thank Kenneth Augustyn for his many considered questions and
comments.  We are indebted to our anonymous reviewer for
motivating us to make explicit some of our core assumptions.

\end{document}